\documentclass
{mn2e}
\usepackage{epsfig}
\usepackage{amsmath, amssymb,bm}

\def \be{\begin{equation}}
\def \ee{\end{equation}}

\def \msun{\rm M_{\odot}}

\begin{document}
\title[QPE or QPO?] {QPE or QPO? -- Quasiperiodic Activity in Low--Mass Galaxy Nuclei}

\author[Andrew King] 
{\parbox{5in}{Andrew King$^{1, 2, 3}$ 
}
\vspace{0.1in} \\ $^1$ School of Physics \& Astronomy, University
of Leicester, Leicester LE1 7RH UK\\ 
$^2$ Astronomical Institute Anton Pannekoek, University of Amsterdam, Science Park 904, NL-1098 XH Amsterdam, The Netherlands \\
$^{3}$ Leiden Observatory, Leiden University, Niels Bohrweg 2, NL-2333 CA Leiden, The Netherlands}

\maketitle

\begin{abstract}
Quasiperiodic eruptions (QPEs) from low--mass galaxy centres may result from accretion from a white dwarf in a very eccentric orbit about the central massive black hole. Evolution under gravitational radiation losses reduces the separation and eccentricity. I note that below a critical eccentricity $e_{\rm crit} \simeq 0.97$, the accretion disc's viscous timescale at pericentre passage is probably longer than the orbital period $P$, and periodic eruptive behaviour is no longer possible. These QPE descendant systems (QPEDs) are then likely to produce quasiperiodic oscillations (QPOs) rather than eruptions, varying more smoothly over the orbital cycle, with duty cycles $\sim1$. I identify 2XMM J123103.2+110648 ($P \simeq 3.9$~hr) and  (more tentatively) RE J1034+396 ($P \simeq 1$~hr) as candidate systems of this type, and find agreement with their deduced eccentricities $e < e_{\rm crit}$.  The absence of eruptions and the lower accretion luminosities resulting from the smaller gravitational radiation losses may make QPED systems harder to discover.
Ultimately they must evolve to have viscous times much longer than the orbital period, and either remain steady, or possibly have infrequent but large outbursts. The latter systems would be massive analogues of the soft X--ray transients produced by low stellar--mass X--ray binaries.

\end{abstract}


\footnotetext[1]{E-mail: ark@astro.le.ac.uk}
\section{Introduction}
\label{intro}
Accreting systems frequently produce quasiperiodic luminosity variations.  Recent observations show that several low--mass galaxy nuclei produce quasiperiodic eruptions (QPEs) in ultrasoft X--rays. Here much of the total accretion luminosity is emitted in short but very bright bursts, which recur at slightly varying intervals. The number of identified QPE systems is likely to grow rapidly, both from new surveys, and from searches through existing databases.

A possible model for these QPE systems (King, 2020; 2022; 2023) invokes a white dwarf (WD) in a very eccentric orbit around the central moderately massive black hole (MBH). This binary system is assumed to originate in a tidal capture (possibly of a stellar--mass binary containing the WD -- see Cufari et al., 2022) which narrowly avoids  completely disrupting the WD. Gravitational radiation (GR) losses shrink the MBH -- WD binary on a timescale of a few thousand years and drive mass transfer. The eruptions occur as the WD fills its tidal lobe at pericentre on each orbit, with some variation, possibly because of angular momentum redistribution within the disc through interaction with resonant disc orbits (King, 2023). 

The restriction to WD orbiters comes from observational selection: these stars fill their lobes at separations small enough that GR drives the orbital evolution and mass transfer at rates providing significant accretion luminosity.  The highly visible nature of the eruptions makes them easily observable. Main sequence (MS)  orbiters must presumably be much more common, as in mature stellar populations the mass in MS stars is typically a factor $\sim 100$ times that in WDs, but do not give bright eruptions because they fill their tidal lobes in wider orbits and have consequently lower GR losses. They nevertheless must collectively contribute to -- and presumably dominate -- the total black hole mass gain in the long term.  This possibility gives QPE systems wider importance in perhaps revealing a new channel for central black hole growth in galaxies.

This picture appears to be quite successful in describing the observed properties of all known QPE systems (including the predicted CNO enhancement in at least one case (Sheng, 2021)), and extending the sample to include other objects not previously recognised as QPE systems, such as HLX--1, with a period $\sim 1$~yr (see King, 2022). 
The current QPE sample now has periods ranging from $\sim 1$~yr down to $\sim 2.4$~hr (King 2022; 2023). For  given observational input (period, black hole mass, and luminosity) the model specifies the companion WD mass and its orbital eccentricity $e$ (see Table 1). In all cases $e$ is very close to unity, and there is a strong tendency for the eccentricity to be less extreme for QPE systems with shorter orbital periods. This is consistent with a population undergoing orbital evolution through GR losses. 

The main point of this paper is to extend this picture. I show that continued GR losses are likely to evolve the binary into a state that ultimately suppresses the eruptions and turns them into rather less noticeable quasiperiodic oscillations (QPOs). I propose two known systems (one rather tentatively) as candidates for these QPE `descendant' systems.

\section{Light Curves}

The quasi--periodic eruptions themselves generally have fast rises and quasi--exponential decays, and duty cycles $< 1$. This is similar in form to soft X--ray transient (SXT) outbursts from stellar--mass X--ray binary systems with low--mass companions, although in SXTs the observed timescale is much longer (weeks to months) than in QPEs. It is now well established that SXT events result from thermal--viscous accretion disc instabilities in which irradiation by the central X--ray source traps the disc in the high accretion--rate state until a large fraction of the disc mass is accreted (King \& Ritter, 1998; see Lasota 2001 for a review). The  similar light curves and strong ultrasoft emission 
seen in QPEs suggests that a similar origin is likely for them too. Provided that the 
system erupts on all or most orbital cycles this means that the disc mass probably does not grow monotonically, explaining why these systems do not seem eventually to have more prolonged outbursts separated by much longer intervals.

This analogy with SXTs raises questions. SXTs are stellar--mass binary systems in which the viscous timescale $t_{\rm visc}$ characterizing the outbursts is far longer than the orbital period. But in QPE systems the reverse holds: the eruptions occupy only a small fraction of the orbital period, which is itself often short compared with those of the SXTs. Evidently this stark difference must be related to the extreme eccentricities $e$ of the QPE sources (see Table 1), which make their accretion discs fundamentally distinct from the circular discs of SXT systems. 

We will see that orbital evolution under GR is likely to lead to a reversal of the 
inequality $t_{\rm visc} < P$ characterizing QPE systems. So to get a full picture of
non--disruptive tidal captures and their possible role in black hole growth, we need to consider these systems, which I denote as QPE descendants (QPEDs). 

Since they have $t_{\rm visc} > P$, QPEDs cannot have eruptive light curves, as the accretion events last longer than the times separating them. The thermal--viscous disc instabilities underlying QPEs must in these systems instead produce quasiperiodic (QPO)  luminosity variations occupying most or all of the orbital cycle (i.e. with duty cycle $\sim 1$). 
But the discussion is inevitably complicated because there are a number of other plausible mechanisms which may produce QPOs. In particular
%
%
%
%
I note that 
there is at least one observed case -- GSN 069 (see Miniutti et al. (2023) and references therein) which shows {\it both} QPEs and QPOs.
%
%
%

In the limit that $t_{\rm visc} >> P$ { in a QPED system} we would expect the variations induced by the orbital motion to become negligible, so that the sources might be observed as steady, or possibly with bright outbursts separated by long intervals 
(i.e. much longer than any plausible orbital period). In this last case they would be high--mass versions of SXTs.

In the rest of this paper I investigate the condition defining the border between QPE and QPED systems, and identify two candidates (one rather tentative) as possible examples of the latter.

\section{The viscous timescale in QPE/QPED systems}
\label{sec:visc}

The accretion discs in QPE sources are evidently able to cool rapidly compared with dynamical timescales, and so are likely to be thin, i.e. have $H << R$, where $H$ is the disc scaleheight at disc radius $R$.
In a circular thin accretion disc the diffusion of surface density $\Sigma$ is well described in term of the viscous timescale 
 \be
 t_{\rm visc} \simeq \frac{1}{\alpha}\left(\frac{R}{H}\right)^2\left(\frac{R^3}{GM}\right)^{1/2},
 \label{tvisc2}
 \ee
where $M$ is the controlling central black hole mass, and $\alpha$ is the standard Shakura--Sunyaev viscosity parameter 
(see e.g. Frank et al., 2002).

In judging the effects on the X--ray light curve in a highly eccentric QPE accretion disc we are particularly interested in the viscous timescale at radii close to the pericentre separation $p = a(1 - e)$, where $a$ is the semi--major axis of the QPE binary. Here the periodic passage of the WD companion 
transfers mass to the accretion disc. In addition it probably also triggers the infall of gas on to the MBH by destabilizing the disc through extraction of angular momentum from resonant disc orbits (King, 2023)\footnote{In contrast, the long--term formation history of the disc is controlled by the longer viscous timescale at large
 disc radii, when the WD is far away from pericentre -- see King, 2022).}.

The dynamical timescale at the disc edge closest to the pericentre passage of the WD is given by setting $R = p = a(1-e)$ in the last factor of (\ref{tvisc2}). Then the viscous timescale there is
\be
t_{\rm visc,\ peri} \simeq \frac{1}{2\pi\alpha}\left(\frac{R}{H}\right)^2(1 - e)^{3/2}P
\label{tperi}
\ee
where $P$ is the orbital period, and I have used Kepler's 2nd law to write 
$a^3 = GMP^2/4\pi^2$. 

Miniutti et al (2023) evaluated this expression numerically for the case of 
GSN~069, assuming $H/R = \alpha = 0.1$, and found that $t_{\rm visc,\ peri} $ (there denoted as $t_{\rm visc}$) is equal to the orbital period $P$ for a critical eccentricity 
\be
e = 0.97
\label{GSN}
\ee
(see the discussion below their Fig. 19). 

This is in fact the universal borderline between QPE and QPED systems, as
\be
t_{\rm visc,\ peri} \simeq  P
\label{cond}
\ee
defines the critical maximum eccentricity $e_{\rm crit}$ for QPED systems,
which gives $e_{\rm crit} = 0.97$
if we assume $H/R \sim \alpha \sim 0.1$ (cf King et al., 2007) in (\ref{tperi}).


Clearly this criterion is not precise, not least because of the uncertainties in estimating $H/R$ and $\alpha$. There is the related question of using $(R^3/GM)^{1/2}$ as the dynamical timescale in regions of the disc where non--circular motions may occur. Since dissipation tends to circularize orbits, we would expect this problem to be less serious in cases where dissipation is strongest -- for example if, as is likely, the disc plane set by the WD orbit is significantly inclined to the black hole spin. Despite the uncertainties, we will see that the rough criterion (\ref{GSN}) appears to be satisfied by the two candidate QPED systems.

An important possible consequence of the inequality $t_{\rm visc,\ peri} < P$ satisfied by QPE systems is that viscous evolution of parts of the disc away from pericentre may produce an eccentric disc. This may
make it possible (cf Lu \& Quataert, 2022) that gas streams from lobe--filling MS stars could intersect the accretion disc near pericentre and give another way of generating QPEs.

\section{QPED systems}

Two systems have appeared in the literature which are candidates for QPED systems: 
2XMM J123103.2+110648 (Lin et al., 2020; Webbe \& Young, 2023) and
RE J1034+396 (Gierli\'nski et al., 2008).
2XMM J123103.2+110648 has significant repetitive variation throughout its observed cycle, and  RE J1034+396 has a near--sinusoidal light curve with a rather coherent period $P \simeq 1$~hr.

One can use the same methods to derive WD masses and orbital 
eccentricities for these two systems as for the QPE systems (i.e. Chen et al., 2022; King 2022). Webbe \& Young (2023) do this for 2XMM J123103.2+110648, but 
remark explicitly that it is difficult to reconcile its observed duty cycle $\sim 1$
with the much shorter duty cycles of the QPE sources.

The interpretation of RE J1034+396 as a QPED system is significantly more problematic than for 2XMM J123103.2+110648.  Its fairly low--mass black hole makes tidal interaction possible, and the derived parameters for an orbiting WD mass donor are encouraging. But
 unlike any other QPE--related system it has a well--developed broad--line region, which may indicate that its accretion disc is more extended, perhaps allowing for direct impacts with the orbiting donor star. In addition, and possibly even more significantly, the QPO in RE J1034+396 dominates the hard power--law X--ray emission and is not detected in the soft X--rays which presumably originate in the accretion disc. Further,  the rms variability is observed to drop (Middleton et al., 2009) as the soft excess starts to dominate the emission, diluting the QPO variability. Evidently more work is needed to see if these differences are compatible with QPED status.

Table 1 below gives the observed and derived properties of these two candidate QPED sources, together with those of the known QPE systems.
The Table shows that both candidate QPED systems 
formally satisfy the critical eccentricity condition (\ref{GSN}), and indeed RE J1034+396 has the smallest eccentricity of any of the sources in Table 1. 

But we also see that
three QPE systems out of the seven known also lie above the line in Table 1, whereas 
I argued above that systems with $t_{\rm visc,\ peri} >P$ (which is equivalent to being above the line) cannot have eruptions. 

One likely possibility here is that this may simply reflect the difficulty in estimating
the fairly variable quantity $(L\Delta t)_{45}$ characterizing the total emitted 
radiation energy per orbit. But other factors such as the black hole spin orientation
may affect the true physical criterion, as suggested in Section \ref{sec:visc} above. 

The fact that GSN 069 shows both QPEs and QPOs suggests another possibility (I am grateful to the anonymous referee for emphasizing this). It is unlikely that the accretion disc is only present when the WD is close to pericentre, and so it presumably evolves viscously during the remainder of the orbital cycle. Since the outer parts of the disc probably lie in the WD's orbital plane this star may interact gravitationally with it. 

One result is that the WD extracts angular momentum from the accretion disc, particularly from resonant gas orbits within it, so limiting its outward spread. This is probably required for mass transfer to be stable (King, 2023). 
But in addition, this interaction may induce precession of the disc, as it is
closely analogous to the superhump interaction found in short--period cataclysmic variables -- see King (2023) for a discussion. Superhumps themselves are a photometric modulation with a period slightly longer than the (circular) orbit, resulting from tidal flexing of the now precessing disc. 

It seems possible that for example the presence of a QPO in the QPE system GSN 069 results from phenomena analogous to superhumps, but now involving a very eccentric binary orbit. One needs to show that it would appear in X--rays, rather than in the optical emission that characterizes superhumps. 

As mentioned in the Introduction, a further possible complication is that in some cases there may be accretion from one or more orbiting main--sequence  
stars in addition to the WD. Another potential source of QPO variability comes from the very likely possibility that the WD orbit -- and therefore the accretion disc -- is inclined wrt the MBH spin. Then the disc may {\it tear} (cf Nixon et al., 2012) into rings which then precess at different frequencies. Raj \& Nixon (2021) investigate this idea. Finally, we should note that the viscous spreading of the disc at locations far from pericentre may allow physical impact of the orbiter with the disc -- Xian et al. (2021) have suggested this as a possible cause of QPEs.

The discussion here and in the previous Section illustrates that both QPEs and QPOs may potentially arise in a number of different ways, particularly depending on the competition between viscous spreading of the disc and the tendency of resonant gravitational interactions with the orbiting WD to inhibit it. 
Investigating this in general requires a full fluid--dynamics treatment of the disc. Despite this,
one reasonably secure inference is that the impact of the mass transfer stream on the accretion disc at pericentre passage does not produce a luminosity which is competitive withe the total accretion luminosity. These are respectively given by $\sim GM\dot M/a(1-e)$ and $\sim GM\dot M/R_g$, and physical consistency requires that the disc radius at pericentre is much larger than the gravitational radius $R_g$. In fact
we see from Table 1 that for all except possibly one of the
QPE/QPED systems we have $R_g \ll a(1-e)$.

\section{Discussion}

We have seen that as GR evolution decreases both $e$ and $P$, QPE systems are likely to evolve into QPED systems, where the eccentricity is below a critical value $e_{\rm crit} \simeq 0.97$ and the accretion disc's viscous timescale is longer than the orbital period.  Then the accretion disc instabilities which made QPEs probably
cause quasiperiodic oscillations (QPOs) instead. In agreement with this,
the two objects identified here as candidate QPED systems both have $e < e_{\rm crit}$ and fairly short orbital periods.

The data of Table 1 do not yet give a clear test of these ideas. Although there is a general tendency for $e$ and $P$ to decrease together, the source properties are defined by three input parameters subject to random and possibly systematic errors, so the sample is currently too small to test this formally. These uncertainties alone may account for the fact that three of the seven QPE sources appear to have eccentricities formally slightly below $e_{\rm crit}$. A larger observational sample would allow one to test for the presence of angular momentum transfer between the orbiting WD and the accretion disc gas, and some of the other physical effects mentioned at the end of the previous Section.

QPED sources may become almost steady once the viscous timescale is much longer than the orbital period, provided that the disc can 
smoothly transport all the mass gained from the WD to the MBH.

If this last condition fails, these systems may have large outbursts recurring at long intervals. A requirement for outbursts is that the effective temperature $T_{\rm irr}$ at the edge of the irradiated accretion disc should in a steady state be less than the hydrogen ionization temperature $T_H \simeq 6500$~K (cf King, Kolb \& Burderi
(1996), who investigate this for the circular discs of stellar--mass SXTs). We use equation (15) of King (2022) to give the accretion rate, and find $T_{\rm irr}$ from
\be
T_{\rm irr}^4 \simeq \frac{\eta c^2(1 - \beta)(-\dot M_2)}{4\pi\sigma R^2}
\left(\frac{{\rm d}H}{{\rm d}R} - \frac{H}{R}\right).
\ee
Here $\eta = 0.1$ is the accretion efficiency, $\beta$ the X--ray albedo, $\sigma$
the Stefan--Boltzmann constant, and $R = a(1-e)$ is the (virtually constant)
disc edge size opposite the WD at pericentre, and the term in large brackets is $\sim H/R \sim 0.1$,
(cf eqn (3) of King, Kolb \& Burderi, 1996). Since $R = a(1-e)$ is virtually constant
until $e$ is $\ll 1$ (the GR losses are effectively a point interaction)
we find that $T_{\rm irr} \gtrsim T_H$ for all likely QPED systems. It appears then that SXT--analogue outbursts of QPED systems are probably rare, but remain a possibility to bear in mind.

Aside from these possibilities, QPE systems seem likely to become less observable as they evolve into QPED objects. This is not a serious restriction on observable lifetimes, as the eccentricity decreases on the GR timescale $- M_2/\dot M_2$, which is of order a few thousand years for all QPE/QPED systems (cf King, 2022).
\section*{DATA AVAILABILITY}
No new data were generated or analysed in support of this research.
\section*{ACKNOWLEDGMENTS}
 It is a pleasure to thank the anonymous referee for very thoughtful and stimulating comments, which led to significant improvements in the paper.

\begin{table*}
\caption{Parameters of the Current QPE/QPED Sample}
\vskip 5.0pt
\centering
{
\setlength{\tabcolsep}{3pt}
{
\hfill{}
  \begin{tabular}{|l||c|c|c|c|c|c|c|c|} 
    \hline
    Source & $P_4$ & $m_5$ & $(L\Delta t)_{45}$ & $m_2$   & $1-e$ & duty cycle
    & light curve
      \\
    \hline \hline
    RE  J1034+396$^a$[?] & 0.37 & 25 & 7.5 & 0.19 & $0.139$ &  1
    &O
    \\
    eRO--QPE2$^b$ & 0.86&  2.5  & 0.8 & 0.16 & $9.0\times 10^{-2}$ & 0.2
    & E
    \\
  XMMSL1 
J024916.6--04124$^c$ & 0.90 &  0.85    & 0.34 & 0.15 & $8.9\times 10^{-2}$ &  0.2
& E
    \\
    RXJ1301.9+2747$^d$ &1.65& 18   &   1.7     & 0.14 & $6.5\times 10^{-2}$ & 0.15
    & E
    \\
    2XMM J123103.2+110648$^e$ & 1.37 & 1& 30 & 0.51 & $3.0\times 10^{-2}$ 
    & 1 &O 
    \\
    \hline
    GSN 069$^f$ & 3.16&  4.0 &  10 & 0.29& $2.5\times 10^{-2}$ & 0.06 & E + O
    \\ 
     eRO--QPE1$^g$ & 6.66 &9.1& 300 & 0.62 & $9.0\times 10^{-3}$& 0.4 & E 
     \\
     ASASSN--14ko$^h$ &937& 700 & $3388$ & 0.56 & 
     $3.7\times 10^{-4}$ & 0.1 &E
     \\
     HLX--1$^i$ & 2000& [10] & 1000 & 0.81 & $1.5\times 10^{-4}$ & 0.3 & E
     \\
     \  . . . & \ . . .  & [0.5]& \ . . .  & 1.43 & $1.2\times 10^{-4}$ & 0.3 & E
     \\
   \hline\hline
  \end{tabular}}
  \hfill{}
  }
  \vskip 0.2truecm
  \begin{itemize}
\item[] 
This table is adapted from Table 1 of King (2023), but now ordered by eccentricity $e$, which decreases upwards (in the direction of GR evolution, which also shortens the orbital periods). 
The systems are characterized by their periods $P_4 = P/10^4 ~{\rm s}$, black hole masses $m_5 = M/10^5\msun$,
and the total energy $(L\Delta t)_{45} = L\Delta t/10^{45}\, {\rm erg}$ emitted per orbit. Assuming these quantities are related by the orbital evolution given by GR emission and the resulting mass accretion,
the method of Chen et al. (2022) then gives the quantities $m_2$ (the WD mass in $\msun$) and $1 - e$, with $e$ the eccentricity of the WD orbit. 
The values of some uncertain parameters have been adjusted slightly, with consequent minor changes in the results for $m_2, 1-e$, compared with Table 1 of King (2023).
The last column on the right specifies the type of light curve, as E or O for 
QPE/QPO. The uncertain status of the QPO system RE J13034+396 as a candidate QPED system is indicated by the [?] sign.
The horizontal line dividing the Table is the rough maximum eccentricity criterion (\ref{GSN}): the two candidate QPED systems defined by duty cycles $\sim 1$ lie above this line, with $e< e_{\rm crit}$, in agreement. { The position of three QPE systems above the line may reflect uncertainties in the input parameters $P_4, m_5, (L\Delta t)_{45}$, or possibly further relevant physics, such as disc breaking,  excitation of resonances within the disc by the WD, or possibly the effects of other MS companion stars.} 

All systems satisfy the criterion 
$m_5(L\Delta t)_{45}^{1/3} \lesssim 10^4$
(King, 2023)
required to avoid the model formally predicting that the WD pericentre distance $a(1-e)$ is smaller than the innermost stable orbital radius, which is itself $\simeq R_g = GM/c^2$, i.e. the black hole's gravitational radius, although ASASSN--14ko is very close to this limit.  King (2023) shows this explicitly for all the sources in the present Table other than the two `O' systems, and it is simple to verify for these directly. 
\\
References to observations: 
$a =$ Gierli\'nski et al. (2008), 
$b =$ Arcodia et al. (2021),
$c = $ Chakraborty et al. (2021), 
$d = $ Giustini et al. (2020),
$e = $ Lin et al. (2020) and Webbe \& Young (2023), 
$f = $ Miniutti et al (2019),
$g = $ Arcodia et al. (2021),
$h = $ Payne et al. (2021),
$i = $ Farrell et al. (2009). Note that there is no secure mass for the black hole in HLX--1, and $m_5 = 0.5$ is the minimum allowing a WD mass respecting the Chandraskhar limit. Assuming a larger $m_5$ gives smaller $m_2$ and $e$.

\end{itemize}
\end{table*}

{}

\end{document}